\documentclass{ifacconf}

\usepackage{graphicx}      
\usepackage{natbib}        
\usepackage{amsmath} 
\usepackage{amssymb}  
\usepackage{amsfonts}
\usepackage{url}

\newtheorem{assumption}{Assumption}
\begin{document}
\begin{frontmatter}

\title{A Passivity-Based Distributed Reference Governor for Constrained Robotic Networks\thanksref{footnoteinfo}} 

\thanks[footnoteinfo]{This work is supported by EM-EASED, FRIA, and JSPS KAKENHI Grant Number JP15H04019. The stay of Tam Nguyen in Tokyo Institute of Technology has been supported by the Erasmus Mundus EASED programme (Grant 2012-5538/004-001) coordinated by CentraleSup\'{e}lec.}

\author[First]{Tam Nguyen} 
\author[Second]{Takeshi Hatanaka}
\author[Second]{Mamoru Doi} 
\author[First]{Emanuele Garone}
\author[Second]{Masayuki Fujita}

\address[First]{Service d'Automatique et d'Analyse des Syst\`{e}mes,
   Universit\'{e} Libre de Bruxelles, Brussels, Belgium (e-mails: tanguyen@ulb.ac.be, egarone@ulb.ac.be)}
\address[Second]{Department of Systems and Control Engineering, 
   Tokyo Institute of Technology, Tokyo, Japan (e-mails: doi@fl.ctrl.titech.ac.jp, hatanaka@ctrl.titech.ac.jp, fujita@ctrl.titech.ac.jp)}

\begin{abstract}                
This paper focuses on a passivity-based distributed reference governor (RG) applied to a pre-stabilized mobile robotic network. The novelty of this paper lies in the method used to solve the RG problem, where a passivity-based distributed optimization scheme is proposed. In particular, the gradient descent method minimizes the global objective function while the dual ascent method maximizes the Hamiltonian. To make the agents converge to the agreed optimal solution, a proportional-integral consensus estimator is used. This paper proves the convergence of the state estimates of the RG to the optimal solution through passivity arguments, considering the physical system static. Then, the effectiveness of the scheme considering the dynamics of the physical system is demonstrated through simulations and experiments.
\end{abstract}

\begin{keyword}
Distributed control and estimation, Control of networks, Mobile robots, Convex optimization, Control under communication constraints
\end{keyword}

\end{frontmatter}

\section{Introduction}

In recent years, teleoperative robotic networks have attracted the interest of several researchers around the world. In bilateral teleoperation, \cite{lee2005bilateral} and \cite{shokri2014comparison} propose control frameworks for the synchronization of bilateral teleoperation systems with communication delays. A number of basic algorithms running on synchronous robotic networks to achieve rendez-vous are analyzed by \cite{martinez2007synchronous2}. The study of motion coordination algorithms for robotic networks is rich in the literature and \cite{bullo2009distributed} summarize the basic tools for coordination algorithms.

However, the main problems of huge telecommunication networks are hardware limitations and communication signal strength, where the latter has been thoroughly analyzed and modeled in \cite{mostofi2010estimation}. \cite{reich2006robot} have shown that the performance of simple and inexpensive onboard hardwares can be similar to the performance of sophisticated and expensive systems that are applied to search and rescue robotic networks. This is why, as a rule of good practice, we usually limit the communication capabilities of each agent and select a small group of leaders that are able to communicate with the teleoperation station.

An important feature to take into account for the control of mobile robotic networks is the ability to manage constraints present in the environment (e.g. walls, holes, obstacles, etc.) and the limitations of the actuators. For geometric constraints, numerous research in the literature for single robot path planning have been carried out using potential fields (\cite{ge2000new}) and grid search (\cite{thorpe1984path}). However, in the case of robotic networks, the importance to manage constraints in an efficient and distributed fashion is brought to light since in general each agent does not share its information to all the agents of the network and the teleoperator is only able to send requests to the leaders. A possible way to deal with the constraints in a distributed way is presented in \cite{soleymani2015distributed}, where a solution based on the reference governor (RG) is proposed.

The RG (\cite{gilbert1994nonlinear}) is an add-on control scheme, which manages the constraints to a pre-stabilized system by suitably changing the applied reference. The first idea to introduce an RG scheme using passivity arguments applied to constrained mobile robotic networks has been proposed in \cite{tam2016passivity}. The proposed set invariance-based RG uses the same passivity arguments for the pre-stabilization of the robotic network, introducing a paradigm for the control of constrained mobile robotic networks. However, the proposed RG solves the problem in a global way but not in a distributed fashion, which will be one of the contribution of this paper.

In theory, passivity has been shown to have close relations with distributed convex optimization. Firstly, \cite{fan2006passivity} developed new decentralized control algorithms that globally stabilize the desired Nash equilibrium by exploiting the passivity property of the feedback loop. Then, the results in \cite{burger2014duality} established a strong and explicit connection between passivity-based cooperative control theory and network optimization theory. The second contribution of this paper is to extend those results to both inequality and equality constraints. The proposed solution is then applied to robotic networks to solve the specific RG problem.

This paper is organized as follows. First, the problem of the mobile robotic network is described. The system is pre-stabilized and then the RG optimization problem is formulated. Then, we focus on a passivity-based distributed optimization scheme to solve the RG problem. We prove that the states of the proposed scheme converge to the optimal solution using passivity arguments for a static robotic network. To demonstrate the effectiveness of the passivity-based distributed RG in real-time, the proposed algorithm is implemented to the pre-stabilized system, where simulations and experiments are carried out.

\section{System Description and Local Feedback}

Consider a system of $n$ mobile robots $\mathcal{V}=\{1,\ldots,n\}$ operating on a plane. The model of the $i$th robot proposed in \cite{de2012controllability} is
\begin{eqnarray}
\dot{q}_i=u_i, & i\in\mathcal{V},
\end{eqnarray}
where $q_i\in\mathbb{R}^2$ and $u_i\in\mathbb{R}^2$ are the position and the velocity input of the $i$th robot, respectively. The robots are able to communicate and the inter-agent communication is modeled by a graph $\mathcal{G}=(\mathcal{V},\mathcal{E}),\,\mathcal{E}\subseteq\mathcal{V}\times\mathcal{V}$. Accordingly, robot $i$ has access to the information of the robots that are belonging to the set of neighbors $\mathcal{N}_i=\{j\in\mathcal{V}|(i,j)\in\mathcal{E}\}$. It is assumed that $\mathcal{V}_h$ is the subset of $\mathcal{V}$ corresponding to the subset of robots that are able to communicate with the teleoperation station, while all the robots in $\mathcal{V} \setminus \mathcal{V}_h$ are able to communicate only with their neighbors. We also introduce the notation $\delta_i$ such that $\delta_i = 1$ if $i\in \mathcal{V}_h$, and $\delta_i = 0$ otherwise.
The system is guided by an operator that sets a reference $r\in\mathbb{R}^2$ to the leaders.
\begin{assumption}
\label{ass:hatanaka}
The graph $\mathcal{G}$ is fixed, undirected and connected.
\end{assumption}

The system dynamics is pre-stabilized through a proportional-integral (PI) consensus estimator (\cite{freeman2006stability}) and a proportional feedback loop. The aggregate system is
\begin{equation}\label{aggregatesystem}
\left[\begin{array}{c}\dot{q}\\\dot{\xi}\end{array}\right]=-\left[\begin{array}{cc}\bar{L} & -\bar{L}\\\bar{L} & 0\end{array}\right]\left[\begin{array}{c}q\\\xi\end{array}\right]+\left[\begin{array}{c}D\otimes I_{2}\\0\end{array}\right]\alpha_r((\textbf{1}_n\otimes I_2)r-q)
\end{equation}
where $q:=[q_1^T,\ldots,q_n^T]^T$ is the coordinate vector, the symbol $\otimes$ is the Kronecker product, $\xi$ the additional integral state, $D\in\mathbb{R}^{n\times n}$ a diagonal matrix whose $(i,i)$-element is equal to $\delta_i$, $\textbf{1}_n$ the n-unit vector, $I_2$ the 2-dimensional identity matrix, $\alpha_r\in\mathbb{R}^+_0$ the gain of the system, and $\bar{L}:=L\otimes I_2$, where $L\in\mathbb{R}^{n\times n}$ is the Graph Laplacian associated to the graph $\mathcal{G}$.

To ensure the formation objective, a pre-defined bias is applied to the robots with respect to the leader. We can then prove the following theorem.

\begin{thm}
The states $q$ of system (\ref{aggregatesystem}) converge to the pre-defined formation reference in absence of constraints.
\end{thm}
\begin{pf}
The proof can be found in \cite{tam2016passivity}.
\end{pf}

The next section formulates the RG optimization problem to deal with the constraints.

\section{Reference Governor Optimization Problem Formulation}

\begin{figure}[ht]
\begin{center}
\includegraphics[width=8cm]{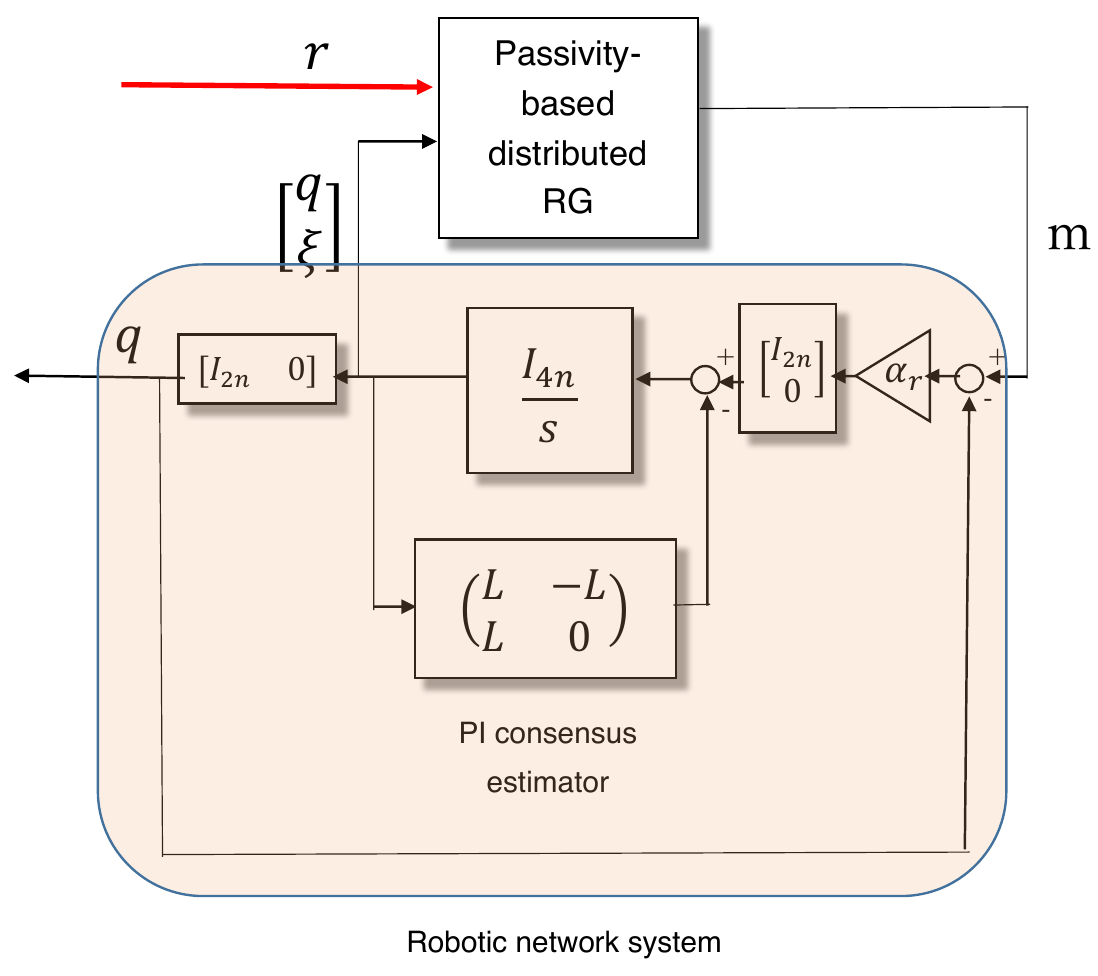}
\caption{Proposed control scheme. The system is first pre-stabilized then augmented with RG for constraints enforcement.}
\label{PBDRGpic}
\end{center}
\end{figure}

Similar to the paradigm proposed in \cite{tam2016passivity}, the constraints of the pre-stabilized system are managed by a set invariance-based RG (\cite{gilbert2001generalized}). The control scheme is depicted in Fig. \ref{PBDRGpic}.

We assume that each robot has sensors detecting obstacles around the robot. Moreover, we assume that the sensors provide a line specifying the boundary of the obstacles detected. The collision-free space is given by the half-space where the robot lies in and this space is denoted by ${\mathcal H}^q_i\subseteq {\mathbb R}^2$. As a consequence, the condition for all robots to avoid collisions with the obstacle detected by robot $i$ is
\begin{equation}\label{hata2}
[q^T \xi^T]^T \in {\mathcal Q}_i := \left\{
[q^T \xi^T]^T|\ q_j \in {\mathcal H}^q_i\ {\forall j\in\mathcal{V}}
\right\}.
\end{equation}
Because the set ${\mathcal Q}_i$ is a polyhedron, it is formulated as
\begin{equation}\label{hata3}
{\mathcal Q}_i := \left\{
[q^T \xi^T]^T|\ A^q_i [q^T \xi^T]^T \leq b^q_i
\right\}
\end{equation}
using some $A^q_i \in {\mathbb R}^{n \times 4n}$ and $b^q_i \in {\mathbb R}^{n}$.
If robot $i$ does not detect any obstacle, then ${\mathcal Q}_i = {\mathbb R}^{4n}$

Let us assume that the input $u_i$ is constrained within a convex polytope ${\mathcal H}^u_i$. Following from (\ref{aggregatesystem}), $u_i$ is given by 
\begin{equation}\label{hata4}
u_i =- (e_i\otimes I_2)(\bar L q - \bar L \xi)+\delta_i (r - q_i),
\end{equation}
where $e_i$ is the $i$-th standard basis of ${\mathbb R}^n$. Therefore, the condition can be formulated as
\begin{eqnarray}
[q^T \xi^T]^T &\in & {\mathcal U}_i(r)
\label{hata5}\\
{\mathcal U}_i(r) &:=& \{
[q^T \xi^T]^T|\  (e_i\otimes I_2)(\bar L q - \bar L \xi)+\delta_i (r - q_i) \in {\mathcal H}^u_i\}\nonumber.
\end{eqnarray}
Since ${\mathcal H}^u_i$ is a convex polytope, 
it is formulated as
\begin{equation}\label{hata6}
{\mathcal U}_i(r) := \left\{
[q^T \xi^T]^T|\ A^u_i [q^T \xi^T]^T \leq b^u_i + B^u_i r
\right\}
\end{equation}
using some $A^u_i \in {\mathbb R}^{\gamma \times 4n}$, $b^u_i \in {\mathbb R}^{\gamma}$
and $B^u_i \in {\mathbb R}^{\gamma \times 2}$, where $\gamma$ corresponds to
the number of conditions specifying ${\mathcal H}^u_i$.

Define 
\[
A_i := \begin{bmatrix}
A_i^q\\
A_i^u
\end{bmatrix},\
b_i := \begin{bmatrix}
b_i^q\\
b_i^u
\end{bmatrix},\
B_i := \begin{bmatrix}
0\\
B_i^u
\end{bmatrix}.
\]
Then, the set ${\mathcal C}_i(r) := {\mathcal Q}_i \cap {\mathcal U}_i(r) $ is given as
\begin{equation}\label{hata7}
{\mathcal C}_i(r) := \left\{
[q^T \xi^T]^T|\ A_i [q^T \xi^T]^T \leq b_i + B_i r
\right\}.
\end{equation}
In the sequel, we assume that the set ${\mathcal C}_i$, namely
$A_i$, $b_i$ and $B_i$,
is a local information of robot $i$ and the other robots do not have access to these information.

Let us now assume that the robotic network modifies the reference $r$
in order to ensure constraints (\ref{hata2}) and (\ref{hata5}), where the modified reference is denoted by $m$. Suppose that a constant signal $m$ is added to (\ref{aggregatesystem})
instead of $r$ and define $q_m = q - (\textbf{1}_n\otimes I_2)m$.
System (\ref{aggregatesystem}) becomes
\begin{equation}\label{hata1}
\left[\begin{array}{c}\dot{q}_m\\\dot{\xi}\end{array}\right]=-\left[\begin{array}{cc}\bar{L}+\alpha_r (D\otimes I_{2}) & -\bar{L}
\\\bar{L} & 0\end{array}\right]\left[\begin{array}{c}q_m\\\xi\end{array}\right]
\end{equation}
Under Assumption \ref{ass:hatanaka}, the matrix $\bar L$ is symmetric and positive semi-definite.
Thus, defining 
\[
V(q_m,\xi) := \frac{1}{2}\|q_m\|^2 + \frac{1}{2}\|\xi\|^2,
\]
it is immediately proved that $\dot V \leq 0$ holds 
along the trajectories of (\ref{hata1}) and hence any 
level set of the function 
is positively invariant for system (\ref{hata1}).
More specifically, define the set
\[
{\mathcal L}_m(c) := \{[q^T\ \xi^T]|\ V(q_m,\xi) \leq c\}
\]
Then, at a time $t$, the state trajectories never get out of the set
${\mathcal L}_m(V(q_m(t), \xi(t)))$ as long as
the constant $m$ will be applied to the system in the future.
Thus, if 
\[
{\mathcal L}_m(V(q_m(t), \xi(t))) \subset {\mathcal C}_i(m),
\]
constraints (\ref{hata2}) and (\ref{hata5}) are never violated. It is worth to note that this set inclusion is a sufficient condition for constraint fulfilment. It is well-known for linear-time invariant systems like (\ref{aggregatesystem}) that, once the constraint sets ${\mathcal C}_i(r)\ (r=1,2,\dots, n)$ are given a priori, a necessary and sufficient condition is provided through offline computation (\cite{gilbert1991linear}).
However, this approach cannot be taken in the present case, 
since these sets are provided online according to the sensing information 
and also no robot can collect all of the sets. 
This is why we take the sufficient condition at the cost of conservatism.

In the next subsections, the global optimization problem is formulated and then the local problem is derived.

\subsection{Global Optimization Problem}

Since the constraints have to be satisfied for all $i$, the global problem to be solved by the robotic network at time $t$ is
\begin{eqnarray}
&&\min_{m\in {\mathbb R}^2}\sum_{i\in {\mathcal V}_h}\|r - m\|^2, \mbox{ subject to}
\label{hata8a}\\
&&{\mathcal L}_m(V(q_m(t), \xi(t))) \subset {\mathcal C}_i(m)\ \ {\forall i}
\label{hata8b}
\end{eqnarray}
Denoting the $l$-th row of $A_i$ and $B_i$ by $A_i^{(l)}$ and $B_i^{(l)}$, respectively, and the $l$-th element of $b_i$ by $b_i^{(l)}$, the constraint ${\mathcal L}_m(V(q_m(t), \xi(t))) \subset {\mathcal C}_i$ in 
(\ref{hata8b}) becomes
\begin{eqnarray}
&&A_i^{(l)}[((\textbf{1}_n\otimes I_2)m)^T\ 0]^T - b_i^{(l)} - B_i^{(l)}m\leq 0 \ {\forall l}
\label{hata9a}\\
&& \frac{1}{\|A_i^{(l)}\|}\left(-A_i^{(l)}\begin{bmatrix}
(\textbf{1}_n\otimes I_2)m\\
0
\end{bmatrix}
+ b_i^{(l)} + B_i^{(l)}m
\right)\geq 
\nonumber\\
&&
\hspace{3cm}
\left\|
\begin{bmatrix}
(\textbf{1}_n\otimes I_2) m\\
0
\end{bmatrix} - \begin{bmatrix}
q(t)\\
\xi(t)
\end{bmatrix}\right\|\ {\forall l}
\label{hata9b}
\end{eqnarray}
At this point, define a function $g_i:{\mathbb R}^2\times {\mathbb R}^{2n}\times {\mathbb R}^{2n}$ such that $g_i(m,q(t),\xi(t))$ coincides with
 (\ref{hata9a})  and  (\ref{hata9b}). 
Remark that the constraint (\ref{hata9a}) is linear in $m$ and hence convex.
The left-hand side of (\ref{hata9b}) is also a convex function in $m$
and the left-hand side is also proved to be convex by showing
positive semi-definiteness of its Hessian.
Thus, the function $g_i(m,q(t),\xi(t))$ is convex in $m$ for any given $q(t)$ and $\xi(t)$ and the problem (\ref{hata8a}) and (\ref{hata8b}) is a convex optimization.

In summary, the problem which the robotic network solves at time $t$ can be
compactly formulated as
\begin{eqnarray}
&&\min_{m}\sum_{i\in {\mathcal V}_h}\|r - m\|^2 \mbox{ subject to:}
\label{hata11a}\\
&&g_i(m,q(t), \xi(t)) \leq 0\ \ {\forall i}
\label{hata11b}
\end{eqnarray}
Following the strategy of the standard reference governors,
the optimal solution to the problem is applied to the system and 
then, at the next time $t+1$, the network again solves the 
problem by replacing $q(t)$ and $\xi(t)$ by new measurements 
$q(t+1)$ and $\xi(t+1)$.
Remark now that since the cost function is strictly convex, 
if there exists an optimal solution for given $q(t)$ and $\xi(t)$,
it must be unique.
However, feasibility of the problem is not always ensured depending on
the set ${\mathcal Q}_i$, namely the locations of the obstacles and of the robots.
Although the issue is basically left to future works, an approach is to expect the human decision to be flexible enough to overcome the problem.

Moreover, another problem is faced and this problem is the main focus of this paper.
Problem (\ref{hata11a}),(\ref{hata11b}) depends
on $A_i$, $b_i$ and $B_i$ for all $i$ but they are local information as mentioned above.
To ensure the information restriction, we need to solve the problem in a distributed fashion.
To this end, we equivalently transform (\ref{hata11a}) and (\ref{hata11b}) into
\begin{eqnarray}
&&\min_{z = (m, z_q, z_{\xi})}\sum_{i\in {\mathcal V}_h}\|r - m\|^2 \mbox{ subject to:}
\label{hata10a}\\
&&g_i(m,z_q, z_{\xi}) \leq 0\ \ {\forall i}
\label{hata10b}\\
&&(e_i\otimes I_2) z_q = q_i(t)\mbox{ and }(e_i\otimes I_2)z_{\xi} = \xi_i(t)\ \ {\forall i}
\label{hata10c}
\end{eqnarray}
From the equivalence to (\ref{hata11a}) and (\ref{hata11b}), if 
(\ref{hata10a})--(\ref{hata10c}) is feasible, the optimal solution to 
(\ref{hata10a})--(\ref{hata10c}) is also unique.
Following the same procedure as above, we can also confirm that 
each element of $g_i$ is convex in $z$ and hence it is also a convex optimization.

In the next subsection, the local optimization problem is derived from the global problem.

\subsection{Local Optimization Problem}

Let us now decompose the problem (\ref{hata10a})--(\ref{hata10c})
into $n$ local problems as follows.
\begin{eqnarray}
&&\min_{z = (m, z_q, z_{\xi})}f_i(z) \mbox{ subject to:}
\label{hata12a}\\
&&g_i(z) \leq 0
\label{hata12b}\\
&&h_i(z) = 0
\label{hata12c}
\end{eqnarray}
where $f_i(z) =\|r - m\|^2$ if $i \in {\mathcal V}_h$ and $f_i(z) = 0$ otherwise,
and $h_i$ is defined so that $h_i(z) = 0$ is equivalent to
$(e_i\otimes I_2) z_q = q_i(t)\mbox{ and }(e_i\otimes I_2)z_{\xi} = \xi_i(t)$.
Then, the local problem consists only of the local information
$A_i$, $b_i$ and $B_i$.
Moreover, it does not depend on the states of other robots.
Thus, if each robot $i$ with the local problem (\ref{hata12a})--(\ref{hata12c})
were able to compute the optimal solution to the global problem
 (\ref{hata10a})--(\ref{hata10c}), they would not need to share the state information among robots. For future developments, denote the k-th element of $g_i$ by $g_{ik}$ and the l-th element of $h_i$ by $h_{il}$. Moreover, define $z^*$ as the optimal solution of  (\ref{hata12a})--(\ref{hata12c}).

In the next section, we present a solution to the above problem.
Although several solutions have already been presented in the literature,
we propose a novel passivity-based solution since it allows one to
integrate other passive components like communication delays
while ensuring the entire system stability
as exemplified in \cite{2016arXiv160904666H}.
A passivity-based distributed optimization algorithm is already presented in \cite{2016arXiv160904666H},
but the equality constraint like (\ref{hata12c}) is not taken into account therein.
We thus extend \cite{2016arXiv160904666H} to problems with equality constraints in order 
to apply the algorithm to the above reference governor problem.  

\section{Distributed Optimization Algorithm and Proof of Convergence for a Static Problem}

In this section, a scheme to solve (\ref{hata12a})--(\ref{hata12c}) is proposed. The proof of the states convergence of the scheme to the optimal solution is provided using passivity arguments when the physical system is static, i.e. $\dot{q}=0$. Then, in the next section, the effectiveness of the proposed solution considering the dynamics of the robotic system is demonstrated through simulations and experiments.

\subsection{Passivity-Based Distributed Optimization Scheme}\label{PBDOsection}

\begin{figure}[ht]
\begin{center}
\includegraphics[width=8.4cm]{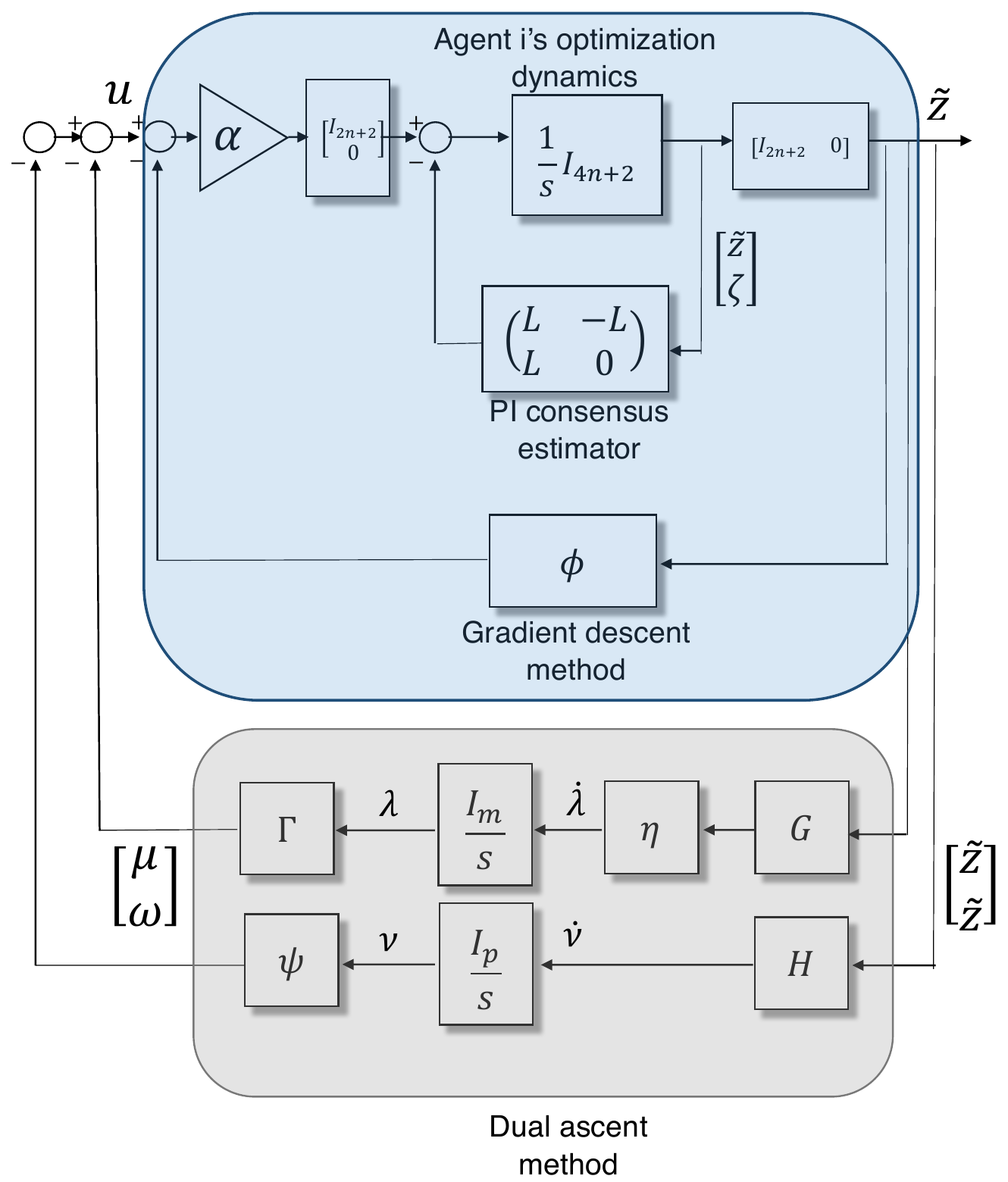}
\caption{Proposed passivity-based distributed optimization architecture, dealing with inequality and equality constraints. $I_{4n+2}$, $I_{m}$, $I_{p}$, $I_n$ are the $4n+2,m,p,n$-dimensional identity matrices, respectively, where $m:=\sum_{i=1}^n m_i$ and $p=:=\sum_{i=1}^n p_i$.}
\label{PBDOarchitecture}
\end{center}
\end{figure}

This subsection gives details about the algorithm to solve (\ref{hata12a})--(\ref{hata12c}). The RG block in Fig. \ref{PBDRGpic} is detailed in Fig. \ref{PBDOarchitecture}, using the state estimate vector $\tilde{z}=[\tilde{z}_1^T,\ldots,\tilde{z}_n^T]^T$, where $\tilde{z}_i$ is the estimate of the optimal solution $z^*$ to (\ref{hata12a})--(\ref{hata12c}) and $\tilde{z}_i\in\mathbb{R}^{(4n+2)}$. For the sake of completeness, consider $m_i\in\mathbb{N}_0$ inequality constraints and $p_i\in\mathbb{N}_0$ equality constraints for agent $i$.

The dual problem associated to (\ref{hata12a})--(\ref{hata12c}) is
\begin{equation}\label{dualproblem}
\begin{aligned}
\text{maximize}&&\sum_{i=1}^n\mathcal{H}_i(\lambda_{i},\nu_{i})\\
\text{subject to}&&\lambda_{i}\geq 0,
\end{aligned}
\end{equation}
where $\mathcal{H}_i(\lambda_i,\nu_i):=\min_{z\in\mathbb{R}^N}\mathcal{L}_i(z,\lambda_i,\nu_i)$ is the Hamiltonian with $\mathcal{L}_i(z,\lambda_i,\nu_i):=f_i(z)+\lambda_i g_i(z)+\nu_i h_i(z)$, and $\lambda_i\in\mathbb{R}^{m_i}$, $\nu_i\in\mathbb{R}^{p_i}$ $(i=1,\ldots,n)$ are the Lagrange multipliers. Define the dual function $\mathcal{H}(\lambda,\nu):=\sum_{i=1}^n\mathcal{H}_i(\lambda_i,\nu_i)$ and the Lagrangian $\mathcal{L}(z,\lambda,\nu):=\sum_{i=1}^n\mathcal{L}_i(z,\lambda_i,\nu_i)$, where $\lambda:=[\lambda_1^T,\ldots,\lambda_n^T]^T$ and $\nu:=[\nu_1^T,\ldots,\nu_n^T]^T$. 

Since strong duality holds, the optimal solution of (\ref{dualproblem}), denoted as $d^*$, satisfies
\begin{equation}
d^*=z^*.
\end{equation}
Moreover, $z^*$ satisfies the Karush-Kuhn-Tucker (KKT) conditions (\cite{boyd2004convex})
\begin{equation}\label{KKT}
\begin{aligned}
\sum_{i=1}^n\nabla f_i(z^*)+\sum_{i=1}^n\lambda_i^*\nabla g_i(z^*)+\sum_{i=1}^n\nu_i^*\nabla h_i(z^*)&=0\\
\lambda_i^*\geq 0,g_i(z^*)\leq 0,\;\;\forall i=1,\ldots,n&\\
\lambda_{ik}^*g_{ik}(z^*)=0,\;\;\forall k=1,\ldots,m_i,\forall i=1,\ldots,n&,
\end{aligned}
\end{equation}
where $\lambda_i^*\in\mathbb{R}^{m_i}$ is the optimal Lagrange multiplier and $\lambda^*_{ik}$ is the k-th element of $\lambda^*_i$.

The main idea to solve (\ref{hata12a})--(\ref{hata12c}) is to minimize the global objective function $f(z)$ through the gradient descent method (\cite{nedic2009distributed}) while maximizing the dual function $\mathcal{H}(\lambda,\nu)$ through the dual ascent method (\cite{boyd2011distributed}). Then, using the PI consensus estimator, the agents will cooperate to converge to the optimal solution.

The next subsection provides the algorithm of the passivity-based distributed optimization scheme.

\subsubsection{Algorithm}\label{algorithmsection}

The algorithm dynamics depicted in Fig. \ref{PBDOarchitecture} is
\begin{equation}\label{algorithm}
\begin{aligned}
\left[\begin{array}{c}\dot{\tilde{z}}\\\dot{\zeta}\end{array}\right]&=-\left[\begin{array}{cc}L & -L\\L & 0\end{array}\right]\left[\begin{array}{c}\tilde{z}\\\zeta\end{array}\right]-\alpha\left[\begin{array}{c}\phi(\tilde{z})\\0\end{array}\right]\\
&-\alpha\left[\begin{array}{c}
\Gamma^T(\tilde{z})\lambda\\0\end{array}\right]-\alpha\left[\begin{array}{c}\Psi^T(\tilde{z})\nu\\0\end{array}\right]
\end{aligned}
\end{equation}
where $\zeta\in\mathbb{R}^N$ is the additional integral variable, $L$ the Graph Laplacian (\cite{ren2008distributed}) associated to the graph $\mathcal{G}$, $\alpha\in\mathbb{R}$ the passivity-based distributed optimization gain, $\phi(\tilde{z}):=[\phi_1(\tilde{z}_1)^T,\ldots,\phi_n(\tilde{z}_n)^T]^T$, where $\phi_i:=\nabla f_i$ is the gradient of the global objective function, and the functions $\Gamma(\tilde{z})$ and $\Psi(\tilde{z})$ are
\begin{equation}
\Gamma(\tilde{z}):=\left(\begin{array}{c c c}
\nabla g_1(\tilde{z}_1)^T & & O\\
&\ddots&\\
& & \nabla g_n(\tilde{z}_n)^T
\end{array}\right),
\end{equation}
\begin{equation}
\Psi(\tilde{z}):=\left(\begin{array}{c c c}
\nabla h_1(\tilde{z}_1)^T & & O\\
&\ddots&\\
& & \nabla h_n(\tilde{z}_n)^T
\end{array}\right).
\end{equation}

The Lagrange multipliers $\lambda,\nu$ are updated as follows. First, define the constraint matrices $G:=[g_1^T,\ldots,g_n^T]^T$ and $H:=[h_1^T,\ldots,h_n^T]^T$. The update algorithm for $\lambda$ is
\begin{equation}\label{updatelambda}
\dot{\lambda}=\eta(\lambda,G),
\end{equation}
where the switch function $\eta(\lambda,G):=\left(\begin{array}{c c c}
\eta_{1,1} & \ldots & \eta_{1,m}\\
\vdots & \ddots & \vdots \\
\eta_{n,1} & \ldots & \eta_{n,m}
\end{array}\right)$ is
\begin{equation}\label{eta}
\begin{aligned}
\eta_{ij}(\lambda_{ij},g_{ij}):=\begin{cases}
0, & \text{if $\lambda_{ij}=0$ and $g_{ij}<0$}\\
g_{ij}, & \text{otherwise},
\end{cases}
\end{aligned}
\end{equation}
where $\lambda_{ij}$ ($i=1,\ldots,n$, $j=1,\ldots,m_i$) denotes the j-th element of $\lambda_i$. This switch block $\eta$ ensures $\lambda_i\geq 0$ (see KKT conditions (\ref{KKT})), where the initial condition must be $\lambda_i(0)\geq 0$.

The update algorithm of $\nu$ is
\begin{equation}\label{updatenu}
\dot{\nu}=h(\tilde{z}).
\end{equation}

\subsection{Convergence to the Optimal Solution}\label{proof}

Define $\tilde{z}^*:=\mathbf{1_n}\otimes z^*$ as the goal state. This subsection will prove the convergence to the optimal solution in three steps:
\begin{enumerate}
\item $\tilde{z}^*$ is a point of equilibrium of the system;
\item the optimization scheme is passive;
\item proof of asymptotic convergence to the optimal solution $\tilde{z}^*$.
\end{enumerate}

\subsubsection{Point of Equilibrium}

We will prove that $\tilde{z}^*$ is a point of equilibrium of the system.

Define the optimal Lagrange multipliers $\lambda^*=[\lambda^{*T}_1,\ldots,\lambda^{*T}_n]^T$ and $\nu^*=[\nu_1^{*T},\ldots,\nu_n^{*T}]^T$. The following lemma holds true.
\begin{lem}\label{pointeq}
There exists $\zeta^*$ such that $(\tilde{z}^*,\zeta^*)$ is an equilibrium of (\ref{algorithm}) for the equilibrium input $\Gamma(\tilde{z}^*)\lambda^*+\Psi(\tilde{z}^*)\nu^*$. In addition, $(\lambda^*,\nu^*)$ is an equilibrium of (\ref{algorithm}) for the equilibrium input $\tilde{z}^*$.
\end{lem}
\begin{pf}
See the proof of Lemma 7 in \cite{2016arXiv160904666H}.
\end{pf}

\subsubsection{Passivity Property}

We will study the passivity property of the passivity-based distributed optimization scheme.

Define $u\in\mathbb{R}^{4n+2}$ as the input of the PI consensus estimator/gradient descent method subsystem as depicted in Fig. \ref{PBDOarchitecture}, $\tilde{z}_c=\tilde{z}-\tilde{z}^*$ as the shifted state estimate, $\tilde{\zeta}:=\zeta-\zeta^*$ as the shifted integral additional state, and $\Delta:=[\mu^T,\omega^T]^T$ as the output of the dual ascent method subsystem, where $[\mu^T,\omega^T]^T:=[\Gamma(\tilde{z})\lambda,\Psi(\tilde{z})\nu]^T$ (see Fig. \ref{PBDOarchitecture}). For the sake of completeness, define the shifted input $\tilde{\Delta}:=[\tilde{\mu}^T,\tilde{\omega}^T]^T$, where $[\tilde{\mu}^T,\tilde{\omega}^T]^T:=[(\mu-\Gamma(\tilde{z}^*)\lambda^*)^T,(\omega-\Psi(\tilde{z}^*)\nu^*)^T]^T$.

First, the passivity of the PI consensus estimator/gradient descent method subsystem is proved (see blue block in Fig. \ref{PBDOarchitecture}).

\begin{lem}\label{lemmapassivity1}
The PI consensus estimator/gradient descent method subsystem is passive from $\tilde{u}:=u-u^*$ to $\tilde{z}_c$ with respect to the storage function $\tilde{S}:=\frac{1}{2}||\tilde{z}_c||^2+\frac{1}{2}||\tilde{\zeta}||^2$, where $u^*:=-\alpha\phi(\tilde{z}^*)$. 
\end{lem}
\begin{pf}
The proof can be found in \cite{hatanaka2015passivity}.
\end{pf}

At this point, the passivity of the dual ascent method subsystem is proved (see gray block in Fig. \ref{PBDOarchitecture}).

\begin{lem}\label{lemmapassivity2}
The dual ascent method subsystem is passive from $\tilde{Z}:=[\tilde{z}_c^T,\tilde{z}_c^T]^T$ to $\tilde{\Delta}$ with respect to the storage function
\begin{eqnarray}\label{storagefunction}
\sum_{i=1}^n U_i, & U_i:=\dfrac{1}{2}\left(||\lambda_i-\lambda_i^*||+||\nu_i-\nu^*_i||\right),
\end{eqnarray}
\end{lem}
where the initial condition for $\lambda_i$ is $\lambda_i(0)\geq 0$.

\begin{pf}
The proof consists in proving that $\sum_{i=1}^n \dot{U}_i\leq \tilde{\Delta}^T \tilde{Z}$.

The first step is to compute the time derivative of $U_i$ in (\ref{storagefunction}), which is
\begin{equation}\label{derivativeUi}
\dot{U}_i=\sum_{j=1}^m\tilde{\lambda}_{ij}\eta_{ij}(\lambda_{ij},g_{ij}(\tilde{z}_i))+\sum_{j=1}^p\tilde{\nu}_{ij}h_{ij}(\tilde{z}_i),
\end{equation}
where $\tilde{\lambda}_{ij}:=\lambda_{ij}-\lambda_{ij}^*$ and $\tilde{\nu}_{ij}:=\nu_{ij}-\nu_{ij}^*$.

The first sum of (\ref{derivativeUi}) is analyzed. The KKT conditions (\ref{KKT}) and the switch block (\ref{eta}) are used to create an inequality. Following from the switch block (\ref{eta}), the terms in the first sum of (\ref{derivativeUi}) becomes
\begin{equation}\label{firstsum}
\begin{aligned}
 \tilde{\lambda}_{ij}\eta_{ij} =
 \begin{cases}
 \tilde{\lambda}_{ij}g_{ij}(\tilde{z}_i)+\lambda^*_{ij}g_{ij}(\tilde{z}_i), & \text{if $\lambda_{ij} = 0$ and $g_{ij}(\tilde{z}_i) < 0$}\\
\tilde{\lambda}_{ij}g_{ij} (\tilde{z}_i), & \text{otherwise}.
\end{cases}
\end{aligned}
\end{equation}
In the first case of (\ref{firstsum}), $g_{ij}(\tilde{z}_i) < 0$ and following from KKT conditions (\ref{KKT}), 
\begin{equation}\label{ineq}
\lambda^*_{ij}g_{ij}\leq 0.
\end{equation}

Following from (\ref{firstsum}),(\ref{ineq}), we deduce
\begin{equation}
\tilde{\lambda}_{ij}\eta_{ij}(\lambda_{ij},g_{ij}(\tilde{z}_i))\leq\tilde{\lambda}_{ij}g_{ij}(\tilde{z}_i).
\end{equation}
As a consequence,
\begin{equation}\label{gradientappear}
\dot{U}_i\leq (\lambda_i-\lambda_i^*)^Tg_i(\tilde{z}_i)+(\nu_i-\nu_i^*)^Th_i(\tilde{z}_i).
\end{equation}

At this point, it is needed to make the gradients appear in the inequality to prove passivity. Using artifices to introduce $g_i(z^*)$ and $h_i(z^*)$ in the inequality, Eq. (\ref{gradientappear}) can be rewritten as
\begin{equation}\label{gradientappear2}
\begin{aligned}
\dot{U}_i\leq&(\lambda_i-\lambda^*_i)^T\{g_i(\tilde{z}_i)-g_i(z^*)\}+(\lambda_i-\lambda^*_i)^Tg_i(z^*)\\
&+(\nu_i-\nu^*_i)^T\{h_i(\tilde{z}_i)-h_i(z^*)\}+(\nu_i-\nu^*_i)^Th_i(z^*).
\end{aligned}
\end{equation}
Note that $(\lambda^*_i)^Tg_i(z^*)=0$ from KKT condition (\ref{KKT}). Since $\lambda_i^T g_i(z^*)\leq 0$ holds, and since the optimal solution $z^*$ satisfies the equality constraint $h_i(z^*)=0$, Eq. (\ref{gradientappear2}) becomes
\begin{equation}
\dot{U}_i \leq (\lambda_i - \lambda^*_i )^T \{g_i(\tilde{z}_i) - g_i(z^*)\}+(\nu_i - \nu^*_i )^T \{h_i(\tilde{z}_i) - h_i(z^*)\},
\end{equation}
which can be rewritten as
\begin{equation}\label{gradientappear3}
\begin{aligned}
\dot{U}_i\leq&\sum_{j=1}^m [\lambda_{ij}\{g_{ij}(\tilde{z}_i) - g_{ij}(z^*)\} - \lambda^*_{ij}\{g_{ij}(\tilde{z}_i) - g_{ij}(z^*)\}] \\
&+ \sum_{j=1}^p [\nu_{ij}\{h_{ij}(\tilde{z}_i) - h_{ij}(z^*)\} - \nu^*_{ij}\{h_{ij}(\tilde{z}_i) - h_{ij}(z^*)\}].
\end{aligned}
\end{equation}

Consider the terms in the first sum of (\ref{gradientappear3}). Because of the convexity of $g_{ij}$, \cite{boyd2004convex} proves that
\begin{equation}\label{gradientappear4}
\begin{aligned}
\begin{cases}
g_{ij} (\tilde{z}_i) - g_{ij} (z^*) &\geq (\nabla g_{ij} (z^*))^T (\tilde{z}_i - z^*),\\
g_{ij}(\tilde{z}_i) - g_{ij}(z^*) &\leq (\nabla g_{ij}(\tilde{z}_i))^T (\tilde{z}_i - z^*).
\end{cases}
\end{aligned}
\end{equation}
Since $h_{ij}$ is affine, $h_{ij}$ is also convex
and $\nabla h_{ij}$ is constant, i.e. 
\begin{equation}
\nabla h_{ij} (z^*)=\nabla h_{ij}(\tilde{z}_i).
\end{equation}
As a consequence, we can deduce that
\begin{equation}\label{gradientappear6}
\begin{aligned}
\begin{cases}
h_{ij}(\tilde{z}_i)-h_{ij}(z^*)&= (\nabla h_{ij} (z^*))^T (\tilde{z}_i - z^*),\\
h_{ij}(\tilde{z}_i) - h_{ij}(z^*) &= (\nabla h_{ij}(\tilde{z}_i))^T (\tilde{z}_i - z^*).
\end{cases}
\end{aligned}
\end{equation}

Therefore, since $\lambda_i\geq 0$ and $\lambda_i^*\geq 0$, using (\ref{gradientappear4}) and (\ref{gradientappear6}) in (\ref{gradientappear3}), the gradients appear as
\begin{equation}\label{gradientappear7}
\begin{aligned}
\dot{U}_i \leq&  \{(\nabla g_i(\tilde{z}_i))^T \lambda_i -(\nabla g_i(z^*)^T)\lambda_i^*\}^T (\tilde{z}_i -z^* )\\
&+ \{(\nabla h_i(\tilde{z}_i))^T \nu_i -(\nabla h_i(z^*)^T)\nu_i^*\}^T (\tilde{z}_i -z^* ).
\end{aligned}
\end{equation}

Therefore, using (\ref{gradientappear7}) in (\ref{storagefunction}), we deduce
\begin{equation}
\begin{aligned}
\sum_{i=1}^n\dot{U}_i&\leq [(\Gamma(\tilde{z})\lambda-\Gamma(\tilde{z}^*)\lambda^*)^T+(\Psi(\tilde{z})\nu-\Psi(\tilde{z}^*)\nu^*)^T]\tilde{z}_c\\
&=\tilde{\mu}\tilde{z}_c+\tilde{\nu}\tilde{z}_c=\tilde{\Delta}^T\tilde{Z},
\end{aligned}
\end{equation}
which concludes the proof.

\end{pf}

In the next subsection, using hybrid Lassale's principle, we can prove asymptotic convergence to the optimal solution $\tilde{z}^*$.

\subsubsection{Convergence}

At this point, we proved that $(\tilde{z}^*,\zeta^*)$ is a point of equilibrium of the system (Lemma \ref{pointeq}) and that the two subsystems (blue and gray blocks in Fig. \ref{PBDOarchitecture}) are passive (Lemmas \ref{lemmapassivity1} and \ref{lemmapassivity2}). Using hybrid Lassale's principle, the following theorem proves that the states converge to the optimal solution $\tilde{z}^*$.

\begin{thm}
Consider the system (\ref{algorithm}),(\ref{updatelambda}),(\ref{updatenu}). The state estimate of the $i$-th agent $\tilde{z}_i$ asymptotically converges to the optimal solution $z^*$ for all $i\in\mathcal{V}$.
\end{thm}
\begin{pf}
%
See the proof of Theorem 3 in \cite{2016arXiv160904666H}.
\end{pf}

In the next section, simulations and experiments are carried out to demonstrate the effectiveness of the passivity-based distributed optimization scheme to solve (\ref{hata12a})--(\ref{hata12c}) in real-time.

\section{Complete Proposed Passivity-based Scheme}

\begin{figure}[ht]
\begin{center}
\includegraphics[width=8.4cm]{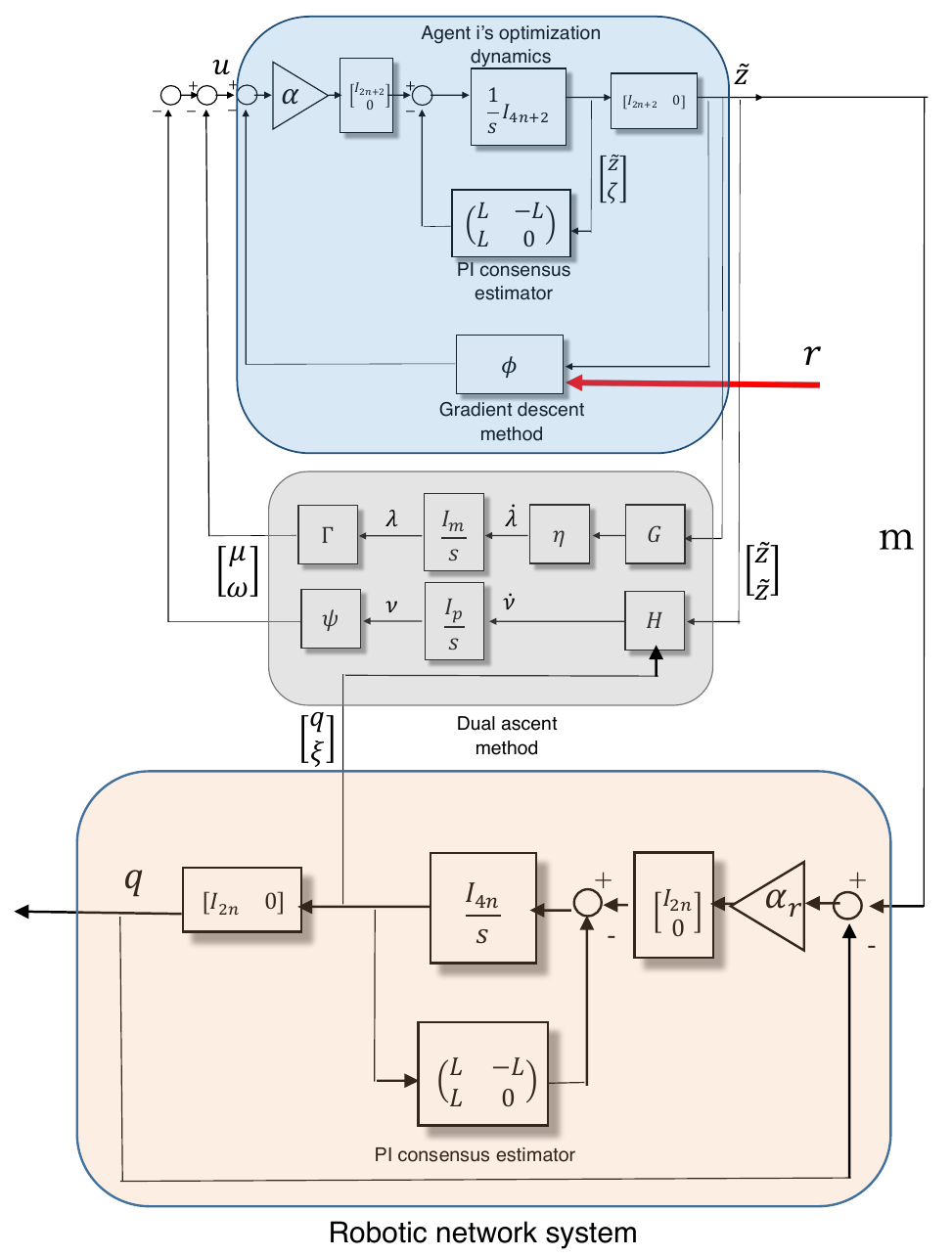}
\caption{Complete scheme considering the dynamics of the robotic network system.}
\label{PBDRGcomplete}
\end{center}
\end{figure}

In this section, the schemes of Fig. \ref{PBDRGpic} and \ref{PBDOarchitecture} are combined as shown in Fig. \ref{PBDRGcomplete}. The proof of the convergence of the entire system is not provided and is let for future works. Instead, the effectiveness of the proposed method in real-time is demonstrated through simulations and experiments.

\subsection{Simulations}\label{simulation}

The system is composed of 5 agents, which communicate in a circle according to the adjacency matrix
\begin{equation}\label{adjmatrix}
A=\left[
\begin{matrix}
0 & 1 & 0 & 0 & 1 \\
1 & 0 & 1 & 0 & 0 \\
0 & 1 & 0 & 1 & 0 \\
0 & 0 & 1 & 0 & 1 \\
1 & 0 & 0 & 1 & 0
\end{matrix}
\right].
\end{equation}
The RG (\ref{algorithm}) is implemented to the pre-stabilized system (\ref{aggregatesystem}).

The leader of the system is agent 5. The formation is added to the system through biases with respect to the leader. The formation is a triangle formation, which is $[-1,1]^T$, $[-1,-1]^T$, $[0,-1]^T$, $[1,-1]^T$ for agents 1, 2, 3 and 4, respectively, biased from agent 5. The initial conditions are $q=[-2,0.7,-1.4,-1,0.2, -1.4, 1.2, -3]^T$, and $\xi$ and all the state estimate variables $\tilde{z}_i$ are initialized at 0. The parameters of the system are $\alpha=2$, $\alpha_r=1$. For simplicity, input constraints are neglected. The line detected by the leader is $x+y=3(m)$.

\begin{figure}[ht]
\begin{center}
\includegraphics[width=7cm]{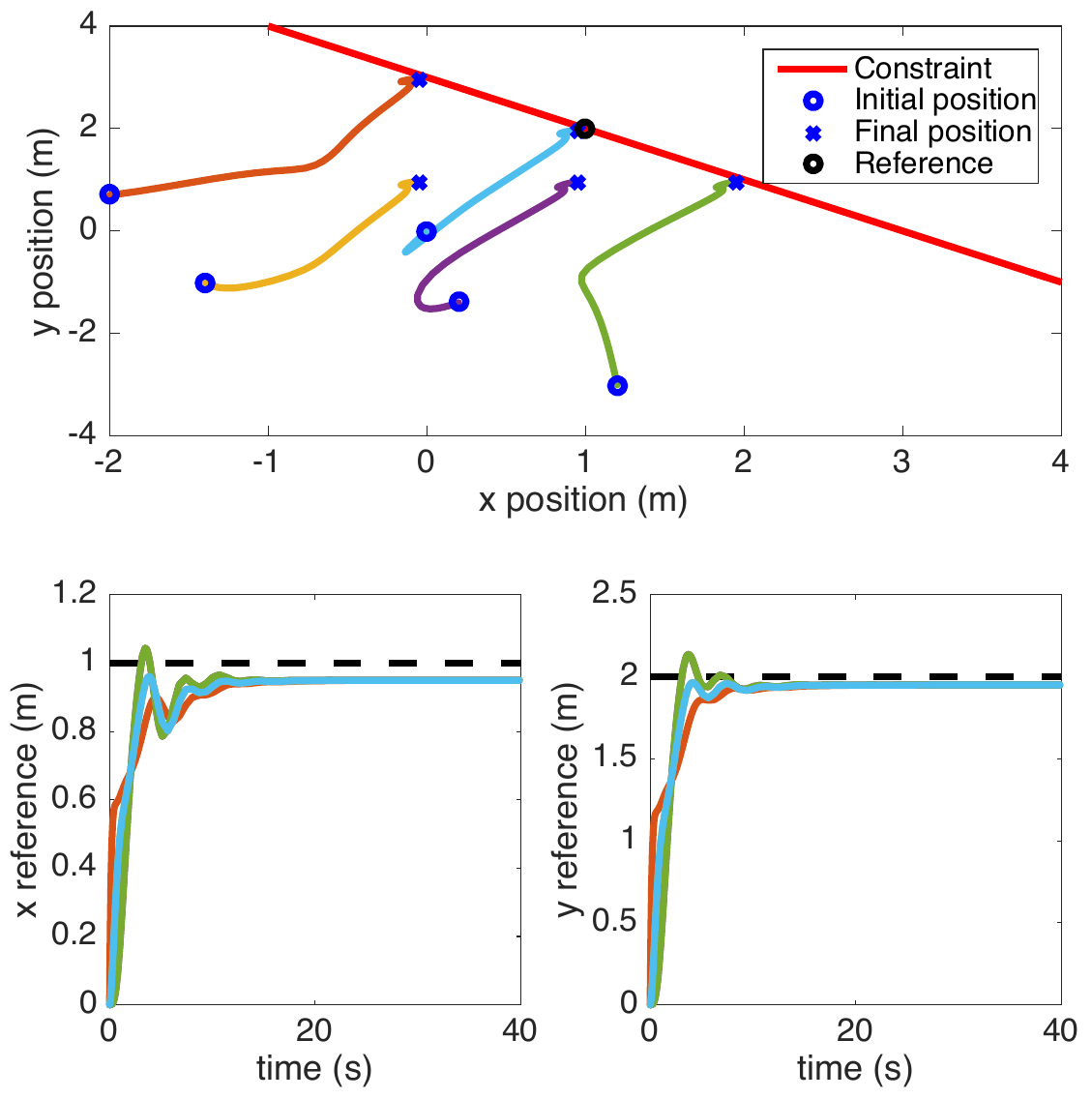}
\caption{Simulations for a reference close to the constraint. Trajectories and modified references of the agents for $r=[1,2]^T$.}
\label{simin}
\end{center}
\end{figure}

\begin{figure}[ht]
\begin{center}
\includegraphics[width=7cm]{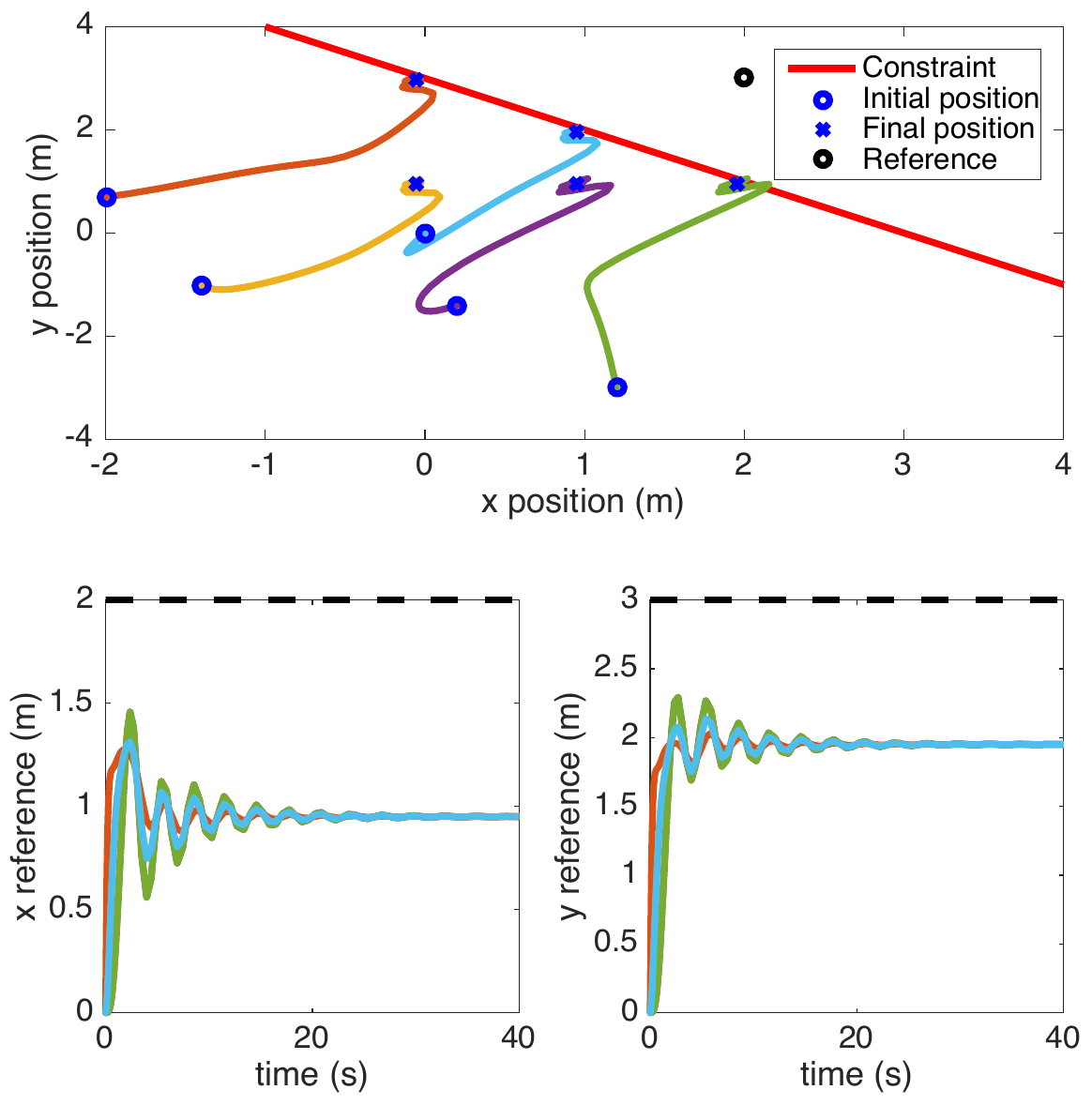}
\caption{Simulations for an inadmissible reference. Trajectories and modified references of the agents for $r=[2,3]^T$.}
\label{simout}
\end{center}
\end{figure}

Consider first a reference chosen close to the constraint $r=[1,2]^T$. The passivity-based distributed RG modifies the reference $r$ so that constraints are enforced during the transients as shown in Fig. \ref{simin}.
Then, consider an inadmissible reference $r=[2,3]^T$. The passivity-based distributed RG modifies the reference $r$ so that the reference remains admissible and ensures constraints satisfaction as shown in Fig. \ref{simout}.

\subsection{Experimental Results}

\begin{figure}[ht]
\begin{center}
\includegraphics[width=7cm]{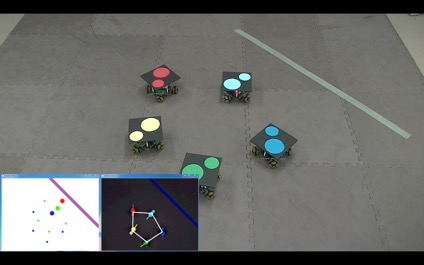}
\caption{Experimental environment captured by cameras.}
\label{experimenttestbed}
\end{center}
\end{figure}

The experimental environment consists of five TDO-48 robots, whose positions are captured by a 30-fps Fire Fly MV camera. The input signal of the robots' motors is a PWM signal, which is sent through a Bluetooth communication using OpenCV 2.4.11. The platform to run the built Simulink program is D-Space 1.1.0.4. Fig. \ref{experimenttestbed} shows the experimental environment.

The agents communicate in a circle according to the adjacency matrix (\ref{adjmatrix}). The RG (\ref{algorithm}) is implemented to the pre-stabilized system (\ref{aggregatesystem}). The leader of the system is agent 5. The formation is added to the system through biases with respect to the leader. The formation is a triangle formation similar to subsection \ref{simulation}. The initial conditions are $\tilde{q}=q(0)$ and $\tilde{r}=q(0)$, where $q(0)$ is the initial real position of the robots, and $\tilde{\xi}$ is initialized at 0. The parameters of the system are $\alpha=7$, $\alpha_r=2.5$. The line detected by the leader is $x+y=3(m)$.

\begin{figure}[ht]
\begin{center}
\includegraphics[width=7cm]{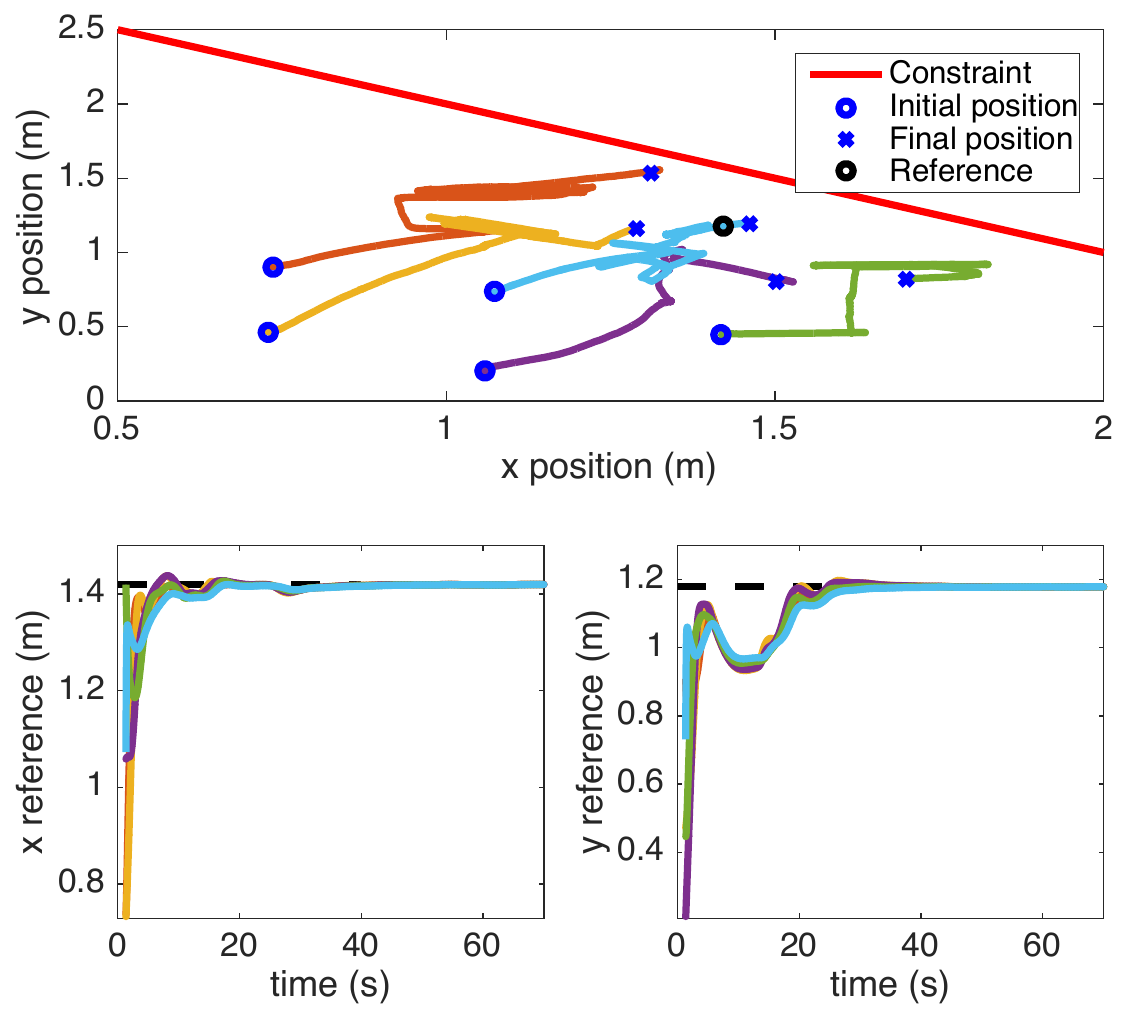}
\caption{Experimental results with a reference close to the constraint. Trajectories and modified references of the agents for $r=[1.4,1.18]^T$.}
\label{expin}
\end{center}
\end{figure}

The first experiment consists in moving the robots to a reference close to the constraint solving (\ref{hata12a})--(\ref{hata12c}) at each time instant. As seen in Fig. \ref{expin}, the trajectories followed by each robot do not violate the constraints at each time instant. Note the small static error due to the difference of static friction that depends on the working area.

\begin{figure}[ht]
\begin{center}
\includegraphics[width=7cm]{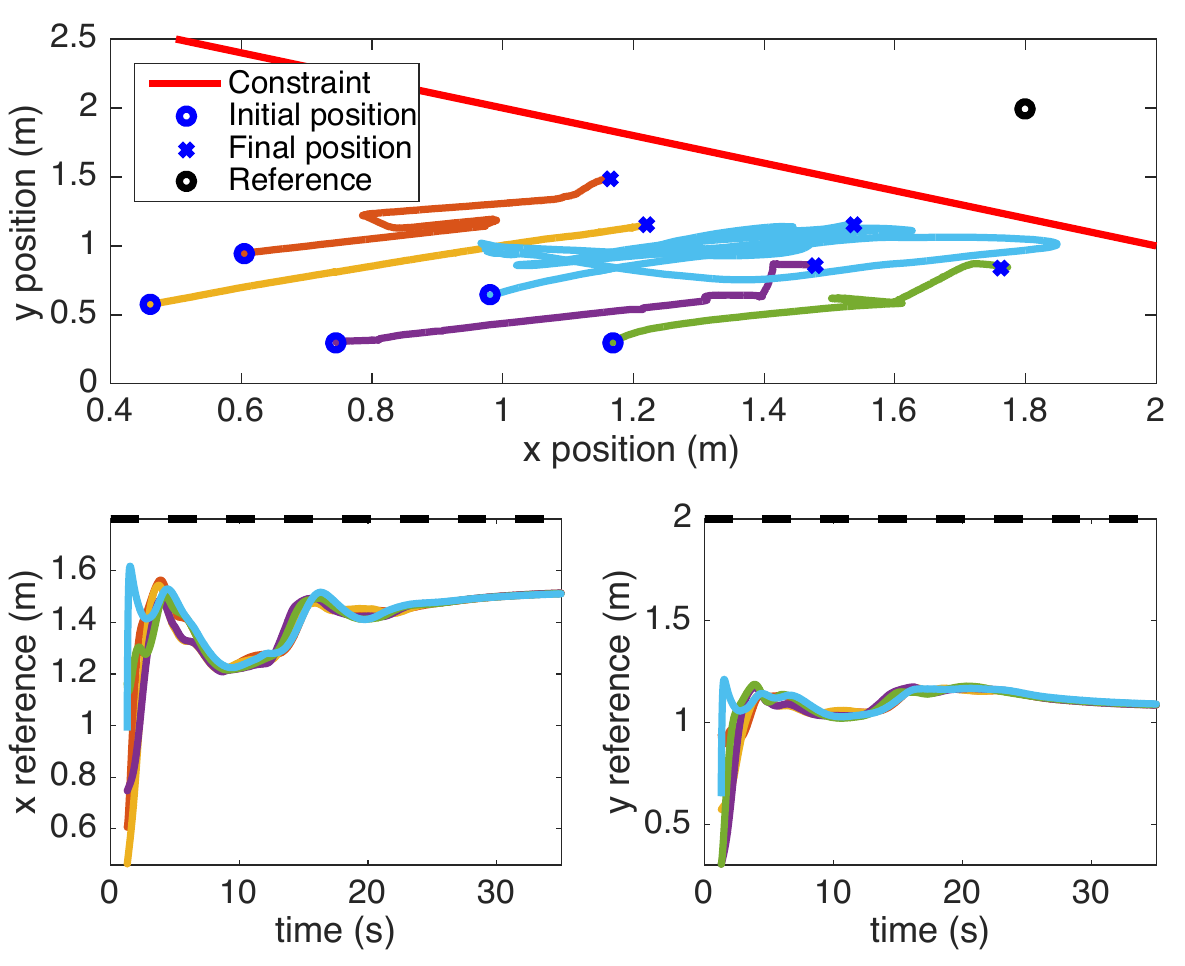}
\caption{Experimental results with an inadmissible reference. Trajectories and modified references of the agents for $r=[1.8,2]^T$.}
\label{expout}
\end{center}
\end{figure}

The second experiment consists in moving the robots to a reference chosen outside the admissible region. As seen in Fig. \ref{expout}, the system behaves as expected and the trajectories followed by each robot do not violate the constraints at each time instant.
The video of the experiments can be found on \url{https://www.youtube.com/watch?v=TeEE9gXt3qQ}.

\section{Conclusions}

A passivity-based distributed RG has been proposed for the control of constrained mobile robotic networks. The distributed RG problem is solved through a passivity-based optimization scheme. This scheme solves a convex optimization problem that is subject to inequality and equality constraints. Passivity arguments are used to prove the convergence of the states to the optimal solution. The main limitation of the passivity-based distributed RG lies in the fact that the problem may be infeasible. In the present paper, the convergence of the interconnected system is demonstrated through simulations and experiments. Future works will aim at proving the convergence of the states of the complete system.


\bibliography{ifacconf}             
                                                   







\end{document}